\documentstyle[12pt]{article}
\newcommand{\be}{\begin{equation}} 
\newcommand{\ee}{\end{equation}} 
\newcommand{\tres}[1]{\buildrel\ldots\over #1} 
\newcommand{\no}{\noindent}
\newcommand{\n}{\label}

\begin{document} 

\title{General behaviour of Bianchi~VI$_0$ solutions with an
exponential-potential scalar field.}

\author{Luis P. Chimento$^{*}$ and Pablo Labraga$^{\dag}$\\
{\small\it $^{(*)}$Departamento de F\'{\i}sica, Facultad
de Ciencias Exactas y Naturales,}\\
{\small\it Universidad de Buenos Aires,
Ciudad Universitaria, Pabell\'{o}n I,}\\
{\small\it 1428 Buenos Aires, Argentina.}\\
{\small\it E-mail: chimento@df.uba.ar}\\
\\
{\small\it $^{(\dag)}$Departamento de F\'\i sica Te\'orica,
Universidad del Pa\'\i s Vasco,}\\
{\small\it Apartado 644, 48080 Bilbao, Spain.}\\
{\small\it E-mail: wtblalop@lg.ehu.es}}

\date{August 4, 1998}

\maketitle

\noindent $^{(*)}$ Fellow of the Consejo Nacional de Investigaciones
Cient\'{\i}ficas y T\'ecnicas.

\baselineskip=18pt

\begin{abstract}

\baselineskip=18pt

The solutions to the Einstein-Klein-Gordon equations without a
cosmological constant are investigated for an exponential potential in a
Bianchi~VI$_0$ metric. There exists a two-parameter family of solutions
which have a power-law inflationary behaviour when the exponent of the
potential, $k$, satisfies $k^2<2$. In addition, there exists a
two-parameter family of singular solutions for all $k^2$ values. A
simple anisotropic exact solution is found to be stable when $2<k^2$.
\vspace{.5cm}

\no KEY WORDS: Bianchi models, exact solutions, asymptotic structure,
power-law inflation.

\end{abstract} 


Inflationary theories claim that the isotropy of the Universe can be
explained by assuming an inflationary expansion in the early universe.
This belief is based on the ``Cosmic no-hair" theorems \cite{wald}.
Heusler~\cite{heusler}, however, proved that by considering the entire
evolution of the scalar field coupled to the gravitational field and
assuming a very large class of reasonable potentials, the evolution of the
anisotropic Bianchi models was similar to that found by Collins and
Hawking \cite{collins}: only those Bianchi models that have as particular
solutions the FRW models isotropize. 

One of the potentials which has received special attention is that of the 
Liouville form. It appears in the Jordan-Brans-Dicke theory and does not
belong to the class of potentials analyzed by Heusler. The asymptotic
behaviour of the Bianchi models with an exponential-potential scalar field
has been studied recently \cite{kitada} using the techniques developed
in \cite{hsu}. When the constant $k$ that gives the slope of the potential
is less than $\sqrt{2}$ then the spacetime inflates and isotropizes.
However, when $k^2>2$, the models that can possibly isotropize are those
of Bianchi types I, V, VII or IX. The isotropization of several scalar
field Bianchi models has been recently studied in \cite{hci} and
\cite{cih}. 

In \cite{afi} the Bianchi~I models with exponential potential were
analyzed reducing the problem to one third order differential equation. In
\cite{ach} a non-local transformation was used to linearize this equation
and the general solution was found. It allows a better insight into the
problem and shows a damped oscillatory behaviour which leads to an
effective negative cosmological constant. Several singular solutions
representing universe models that have either anisotropic or isotropic
Friedmann--Robertson--Walker final stages were found. There are also
solutions which avoid the initial singularity and others with a finite
time span. In this paper we shall extend this analysis to the
Bianchi~VI$_0$ type, investigating the behaviour of the solutions near the
singularity and its asymptotic stability in the far future. 


\vspace{1cm}

The metric corresponding to a Bianchi~VI$_0$ cosmological model can be
written in the following form:

\be \n{1} ds^2=e^{f(t)}\left( -dt^2+dz^2\right) + G(t)\left(
e^zdx^2+e^{-z}dy^2\right) . \ee

\no The pertinent Klein-Gordon and Einstein field equations for the
metric~(\ref{1}) are as follows:

\be \n{3} \ddot\phi +\frac{\dot G}{G}\dot\phi + e^f\frac{\partial V}
{\partial\phi}=0, \ee

\be 
\n{4}
\frac{\ddot G}{G}=2e^fV, 
\ee 

\be 
\n{5}
\frac{\ddot G}{G}-\frac{1}{2}\frac{\dot G^2}{G^2}-\dot f
\frac{\dot G}{G}+\frac{1}{2}+\dot\phi^2=0. 
\ee 

\no We will consider a homogeneous self-interacting scalar field
($\phi=\phi(t)$) with an exponential potential $V=\Lambda e^{k\phi}$
. As can be easily shown

\be
\n{9}
\dot\phi=\frac{m}{G}-\frac{k}{2}\frac{\dot G}{G},
\ee

\no is a first integral of the Klein-Gordon equation and $m$ is an
arbitrary integration constant. From Eqs.(\ref{4}-\ref{9}), we get

\be 
\n{11}
\ddot G^2 G-K\ddot G\dot G^2-\tres G\dot G G+\frac{1}{2}\ddot G G^2+
m^2\ddot G=0;  \qquad K=\frac{k^2}{4}-\frac{1}{2},
\ee

\no First we will consider the special case in which $m^2=0$, and after
that, we will pay attention to the most general case $m^2\ne 0$.

\vspace{1cm}

\no {\it $m^2=0$ case} 
\vspace{.5cm}

\no To analyze the asymptotical behaviour of the solutions we define the
new variables $h=\frac{\dot G}{G}$ and $d\eta=hdt=d{\ln{G}}$, so that
Eq.(\ref{11}) now reads

\be
\label{primes}
h^2 h'' + (1+K) h^2 h' + Kh^3 - \frac{1}{2} (h' + h)=0,
\ee

\no with $h'=\frac{dh}{d\eta}$. The simplest solution of
equation~(\ref{primes}) is $h^2=\frac{1}{2K}$ for $K>0$. Hence, we get

\be
\n{13.1}
G=G_0e^{\sqrt\frac{1}{2K}\,t},
\ee

\no while the remaining component of the metric $f$ and the scalar field
$\phi$ can be obtained from Eqs.(\ref{4}) and (\ref{9}) respectively.


Now, we assume that the function $G$ vanishes or diverges at $t=0$ and its
leading term (this assumption is justified in Ref.\cite{luis1}), is given by

\be
\n{14}
G(t)\simeq G_0\,t^n.
\ee

\no Substituting this expression into Eq.(\ref{primes}), we can see that
the last two terms are negligible compared with the first four terms. So

\be
\n{15}
\frac{1}{t^4}\left[ n^2-(1+K)n^3+Kn^4\right]\simeq 0
\ee

\no is the equation that determines the values of the exponent $n$.
Inserting this value of $n$ and Eq.(\ref{14}) in Eqs.(\ref{4}) and
(\ref{9}) , we obtain the approximate scalar field and metric in the limit
$t\rightarrow 0$. 

\no For $1<K$ which corresponds to values of $6<k^2$, with $0<n<1$, we
have $\Lambda <0$ for Eq.(\ref{4}), so the potential must be negative.
However, it is positive for $K<1$. 

\no For $-\frac{1}{2}<K<0$ which corresponds to values of $k^2<2$, with
$-\infty<n<-2$, the metric coeficients $G$ and $e^f$ diverge when
$t\rightarrow 0$. 

The next term in the expansion of the function $G$, can be obtained
solving Eq.(\ref{primes}) when the last two terms are neglected. Its
general implicit solution is

\be
\n{19.1}
G=\left[-\frac{C_1}{C_2}+C_3e^{(1-K)C_2\tau}\right]^{\frac{1}{1-K}},
\qquad  K\ne 1,
\ee

\no where $\tau=\int{G^{-K}\,dt}$.

\no For $-\frac{1}{2}<K<0$ and $1<K$ the function $G$ diverges (vanishes)
in the limit $\tau\to\pm\infty$ according to $C_2>0$ or $C_2<0$. In this
case we get

\be
\n{19.4}
G(t)\simeq at+b|t|^{\frac{1}{K}},
\ee

\no There exists a two-parameter family of solutions that behaves as
$G\simeq|t|^{\frac{1}{K}}$. Also, $G$ vanishes at a finite time $\tau_0$
for $K<1$, provided that $C_1C_2C_3>0$. Its approximate expansion is

\be
\n{19.5}
G\simeq at+b|t|^{2-K}.
\ee

\no Thus a two-parameter family of solutions behaves as $G\simeq t$.


The behaviour and stability of the solutions can be investigated by writing
Eq.(\ref{primes}) as the equation of motion for a dissipative or
antidissipative system,

\be 
\n{20}
\frac{d}{d\eta}\left[\frac{h'^2}{2}+K\frac{h^2}{2}-\frac{1}{2}\ln{h}
\right]=-\left[-\frac{1}{2h^2}+(1+K)\right] h'^2,
\ee 

\no with the potential

\be 
\n{21}
{\cal V}(h)=K\frac{h^2}{2}-\frac{1}{2}\ln{h}.
\ee 

\no Equation~(\ref{20}) presents local minima when $h_0^2=\frac{1}{2K}$,
for $K>0$ (i.e. $k^2>2$). As the dissipative term given by the right-hand
side of Eq.(\ref{20}) is negative definite in the asymptotic regime, the
bracket on the l.h.s. of Eq.(\ref{20}) defines a Liapunov Function
\cite{cesari},\cite{luis},\cite{ale}. Then, the exact solution
(\ref{13.1}) is stable for $t\rightarrow \infty$ and for any initial
condition. Studying the behaviour of the solutions around these
equilibrium points we have: 

\no For $K>0$ ($k^2>2$) there is a two-parameter family of stable
solutions that behaves as (\ref{13.1}). When $K>\frac{1}{8}$ it can be
shown that the solutions cut the $h$ axis in the phase plane $(h,\dot h)$
an infinite number of times. Therefore they spiral in the phase plane
around the constant solution $h_0$. For $-\frac{1}{2}<K\le \frac{1}{8}$
the solutions do not cut the $h$ axis or they cut it once. 


\vspace{1cm}

\no {\it $m^2\ne 0$ case} 
\vspace{.5cm}

\no We are going to proceed now in the same manner as we did in the
previous case. In the limit $t\rightarrow 0$, Eq.(\ref{11}) with 
$m^2\neq 0$ gives

\be 
\label{enes} 
\frac{1}{t^4}\left[n^2-(1+K)n^3+Kn^4\right]-\left(\frac{1}{2}+
\frac{m^2}{t^{2n}}\right)\frac{1}{t^2}\left[ n^2-n\right]=0.
\ee 

\no From this equation we see that the term which comes from $\ddot GG^2$
in (\ref{11}) can be neglected and the remaining approximate equation can
be solved using the method described in \cite{ch}. In this case, we obtain

\be
\n{11.2}
G=\left[\mbox{e}^{-\theta/2}\left(C_1\mbox{e}^{\lambda\theta}+
C_2\mbox{e}^{-\lambda\theta}\right)\right]^{\frac{1}{K-1}}, \qquad K\ne 1,
\ee

\no where

\be
\n{11.6}
\lambda=\frac{\left[1-4\beta\right]^{1/2}}{2}, \qquad \beta=(K-1)
\frac{m^2}{C^2},
\ee

\no and $\theta=-C\int{G^{-1} d\,t}$.

\no $\beta\le 1/4$: For $-\frac{1}{2}<K<0$ and $1<K$ the coefficient of
the metric $G$ diverges (vanishes) at a finite time $\theta=\theta_0$ when
$\mbox{sgn\,}C_1\ne \mbox{sgn\,}C_2$. In this case, we get

\be
\n{11.8}
G\simeq at+b|t|^{1/K} \qquad\mbox{for} \qquad -\frac{1}{2}<K<0
\qquad\mbox{and}  \qquad 1<K,
\ee

\no This two-parameter family of solutions behaves as $|t|^{1/K}$ when
$t\to0$.  Also, $G$ vanishes in the limit $\theta \to\pm\infty$ for $K<1$
and its expansion is 

\be 
\n{11.7} 
G^\pm\,\simeq a^\pm\,t+b|t|^{2-K+\frac{m^2}{(a^\pm)^2}}
\qquad\mbox{for} \qquad K<K_0=1+\frac{m^2}{(a^\pm)^2}, 
\ee

\no where $a^\pm\,=C\left[1/2\pm\,\lambda\right]/(K-1)$.

\no $\beta>1/4$: Eq.(\ref{11.2}) describes damped oscillatory solutions. 
These solutions are compatible only with an effective negative
cosmological constant \cite{ach}. 


In the general case Eq.(\ref{primes}) can be written as the equation
of motion for a dissipative or antidissipative system as follows
\be
\label{liapunov} 
\frac{d}{d\eta}\left[ \frac{h'^2}{2}+{\cal V}(h)\right] = 
2m^2e^{-2\eta}\ln{h}-\left\{(1+K)-\frac{1+2m^2e^{-
2\eta}}{2h^2}\right\}h'^2,
\ee 

\no where now the ``potential" is

\be 
\n{30} 
{\cal V}(h)=\frac{Kh^2}{2}-\left\{
\frac{1}{2}+m^2e^{-2\eta}\right\} \ln{h}, 
\ee

\no and the local extreme points will be given by

\be
\n{31}
h_0^2=\frac{1+2m^2e^{-2\eta}}{2K}.
\ee

\no Requiring that $K>0$, the extreme points are actually minima and the
right-hand side of Eq.(\ref{liapunov}) is negative definite, so that the
bracket on the l.h.s of Eq.(\ref{liapunov}) defines a Liapunov function.
Near the equilibrium points the approximate equation governing the
trajectories on the phase plane, for $t\rightarrow\infty$, is given by
Eq.(\ref{31}) whose solution is

\be 
\n{33}
G_{min}=\pm\left\{-\frac{m^2}{2}e^{\sqrt{\frac{1}{2K}}\,(t-t_0)} +
e^{-\sqrt{\frac{1}{2K}}\,(t-t_0)}\right\} ,
\ee 

\no showing that the final anisotropic solution is asymptotically stable
and coincides with the solution~(\ref{13.1}) . For $k^2>2$ the late-time
behaviour of an inhomogeneization of a Bianchi I model \cite{afi2} is like
that of Bianchi VI$_0$. 

The limit behaviour of the scalar curvature when $t\to 0$ has two
different possibilities depending on the values of $K$: 

\be
\n{RG}
R\sim\left\{
\begin{array}{lcc}
t^{-\left[2+\frac{1}{K}\right]} &
\hbox{for} & -\frac{1}{2}<K<0\quad\hbox{and}\quad 1<K, \\
&&\\
t^{-\left[\frac{3}{2}+\left(\frac{m}{a^\pm}-\frac{k}{2}\right)^2\right]
} & \hbox{for} &  0<K<K_0.
\end{array}
\right.
\ee

\no We conclude that there exists a two-parameter family of singular
solutions which describe a universe that begins from (or ends in) a
singularity for any values of the constants $m$, $C$ and $k^2$. For these
solutions the scalar curvature diverges when $t\to 0$. However, for
$-\frac{1}{2}<K<0$ it vanishes asymptotically and the solution remains
regular. 

We can also see that the shear-expansion ratio,

\be
\n{final}
\frac{\sigma}{\Theta}=\frac{\sqrt{6}}{3} \frac{\dot f G - 
\dot G}{\dot f G + 2\dot G},
\ee

\no which is usually considered as an indicator of the isotropization, does
not vanish for the asymptotical solution given by (\ref{13.1}). 
Therefore, we can say that none of the solutions with $k^2>2$, for which
that study was valid, isotropize. 

\vskip 1cm

Some results of this paper are related to those ones obtained in
\cite{afi}, \cite{ach} and \cite{afi2}. Solutions with $k^2<2$ show a
power-law inflationary behaviour\cite{lm} while solutions with $k^2>2$ do
not inflate or isotropize, since Bianchi~VI models do not have FRW as
particular solutions \cite{collins}. 

\vspace{1cm}
{\large\bf Acknowledgements}
\vspace{.5cm}

The authors wish to thank Prof. J.~Ib\'a\~nez and Prof. A.~Feinstein for
very helpful discussions about this problem. One of us (P.L.) wants to
acknowledge the hospitality of the Department of Physics at the University
of Buenos Aires, Argentina. P.L.'s work was supported by the Basque
Government fellowship B.F.I.92/090. This work was supported by the Spanish
Ministry of Education Grant (CICYT) No PB93-0507 and by the University of
Buenos Aires Grant EX-260. 



\begin{thebibliography}{20}

\bibitem{wald}
R.~M.~Wald
{\it Phys. Rev. D \/} {\bf 28} (1983) 2118;\\
L.~G.~Jensen and J.~A.~Stein-Schabes
{\it Phys. Rev. D \/} {\bf 35} (1987) 1146.

\bibitem{heusler}
M.~Heusler
{\it Phys. Lett. B \/} {\bf 253} (1991) 33.

\bibitem{collins}
C.~B.~Collins and S.~W.~Hawking
{\it Astrophys. J. \/} {\bf 180} (1973) 317.

\bibitem{kitada}
Y.~Kitada and K.~Maeda
{\it Phys. Rev. D \/} {\bf 45} (1992) 1416;\\
J.~Ib\'a\~nez, R.~J.~van den Hoogen and A.~A.~Coley
{\it Phys. Rev. D \/} {\bf 51} (1995) 928.

\bibitem{hsu}
L.~Hsu and J.~Wainwright
{\it Class. Quantum Grav. \/} {\bf 3} (1986) 1103;\\
J.~Wainwright and L.~Hsu
{\it Class. Quantum Grav. \/} {\bf 6} (1989) 1409.

\bibitem{hci}
R.~J.~van den Hoogen, A.~A.~Coley and J.~Ib\'a\~nez,
{\it Phys. Rev. D \/} {\bf 55} (1997) 5215.

\bibitem{cih}
A.~A.~Coley, J.~Ib\'a\~nez and R.~J.~van den Hoogen, 
{\it J. Math. Phys.\/} {\bf 38} (1997) 5256.

\bibitem{afi}
J.~M.~Aguirregabiria, A.~Feinstein and J.~Ib\'a\~nez
{\it Phys. Rev. D \/} {\bf 48} (1993) 4662.

\bibitem{ach}
J.~M.~Aguirregabiria and L.~P.~Chimento
{\it Class. Quantum Grav. \/} {\bf 13} (1996) 3197.

\bibitem{afi2}
J.~M.~Aguirregabiria, A.~Feinstein and J.~Ib\'a\~nez
{\it Phys. Rev. D \/} {\bf 48} (1993) 4669.

\bibitem{luis1}
L.~P.~Chimento
{\it Class. Quantum Grav. \/} {\bf 5} (1988) 1137.

\bibitem{cesari}
L.~Cesari
{\it Asymptotic Behaviour and Stability Problems in Ordinary Differential
Equations \/}(new York: Academic Press) (1963)

\bibitem{luis}
L.~P.~Chimento
{\it Class. Quantum Grav. \/} {\bf 6} (1989) 1285.

\bibitem{ale}
L.~P.~Chimento and A.~S.~Jakubi,
{\it Class. Quantum Grav. \/} {\bf 10} (1993) 2047.

\bibitem{ch}
L.~P.~Chimento,
{\it J. Math. Phys.\/} {\bf 38} (1997) 2565.

\bibitem{lm}
F.~Lucchin and S.~Matarrese
{\it Phys. Rev. D \/} {\bf 32} (1985) 1316.

\end{thebibliography}
\end{document}